\documentclass[12pt]{article}
\usepackage{psfig}

\newcommand{\be}{\begin{equation}}
\newcommand{\ee}{\end{equation}}
\newcommand{\bea}{\begin{eqnarray}}
\newcommand{\eea}{\end{eqnarray}}
\def\Journal#1#2#3#4{{#1 {\bf #2}, #3 (#4)}}

\makeindex
\begin{document}

\title{Nonanalyticity of the beta-function and systematic errors in
field-theoretic calculations of critical quantities}

\author{Michele Caselle \\
Dip. di Fisica dell'Universit\`a di Torino and INFN, Sez. Torino,\\
I-10125 Torino, Italy \\[2mm]
Andrea Pelissetto \\
Dip. di Fisica dell'Universit\`a di Roma ``La Sapienza" and 
INFN, Sez. Roma 1,\\  I-00185 Roma, Italy \\[2mm]
Ettore Vicari \\
Dip. di Fisica dell'Universit\`a di Pisa and INFN, Sez. Pisa,\\
I-56100 Pisa, Italy }

\maketitle

\begin{abstract}
We consider the fixed-dimension perturbative expansion. We discuss the 
nonanalyticity of the renormalization-group functions at the fixed point
and its consequences for the numerical determination of critical quantities.
\end{abstract}

\index{Critical phenomena}
\index{Perturbative expansions}
\index{Ising model}
\index{Critical exponents}
\index{High-temperature expansions}

In the last thirty years there has been a significant progress in the 
understanding of critical phenomena. It has been realized that the 
behavior in the neighborhood of a critical phase transition,
i.e. a transition characterized by long-range correlations,
is determined by very few properties,
the space dimensionality, the range of the interactions, the number of 
components of the order parameter, and the symmetry of the Hamiltonian.
This means that physically different systems may have the same critical 
behavior.  For instance, a simple fluid
at the liquid-vapor transition and a uniaxial magnet at the Curie point behave 
identically: Critical exponents, dimensionless amplitude ratios,
scaling functions are numerically equal.
This phenomenon, which is referred to as universality, has been 
understood within the Wilson's renormalization-group (RG) approach. 
The conceptual setting is thus quite well established,
and the theory of critical phenomena has thus reached the maturity 
of well-verified theories like, for instance, QED or the 
standard model of weak interactions. Nonetheless, it is important to improve 
experiments and theoretical calculations in order to understand the limits 
of validity of these theories. In order to test QED and the standard model, 
several experiments have provided accurate estimates that can be 
directly compared with the theoretical predictions. The most classical ones
are the experiments on the $g$-factor of electrons and muons, and on the 
Lamb shift in hydrogen. In the theory of critical phenomena, the superfluid
transition in ${}^4$He plays a very special role, since it is essentially
the only case in which one can determine a critical exponent with an 
accuracy of $10^{-4}$. This is due to a combination of particularly 
favorable conditions: the singularity in the compressibility of the 
fluid is particularly weak; it is possible to prepare very pure 
samples; experiments may be performed in a microgravity 
environment (on the Space Shuttle, for instance), thereby reducing 
the gravity-induced broadening of the transition. A recent 
experiment \cite{LSNCI-96,errata-00} 
obtained an extremely accurate estimate\footnote{The original 
result reported in Ref. \cite{LSNCI-96} was incorrect. The new estimate
is reported in Ref. \cite{errata-00}. The error is a private communication
quoted in Ref. \cite{CHPRV}.} of $\alpha$,
$\alpha = - 0.01056(38)$. This result should be compared with 
the most precise theoretical estimates:
The analysis of high-temperature (HT) 
expansions gives\protect \cite{CHPRV,CPRV-00} 
$-$0.0146(8), $-$0.0150(17); 
Monte Carlo (MC) simulations give\protect \cite{CHPRV,HT-99}   
$-$0.0148(15), $-$0.0169(33); 
The $d=3$ perturbative expansion gives\protect \cite{JK-00,GZ-98} 
$-$0.0112(21), $-$0.011(4).
There is a clear discrepancy between the most accurate theoretical estimates 
and the experimental result. However, in order to understand 
if the difference is truly significant, we must ask the 
question: `Are the quoted errors reliable?'
Our experience, looking backward in time, is that there is a natural 
tendency to be overconfindent in own's results, and thus to systematically
underestimate the errors: As Hagen Kleinert \cite{JK-00} put it, each one has 
a tendency to apply the ``principle of maximal optimism."  Clearly,
further theoretical and experimental investigation is needed 
to settle the problem.

In order to set correct error bars, it is necessary, although clearly 
not sufficient, to have a good understanding of the possible sources of 
systematic error. In MC and HT works, 
most of the 
systematic error is due to the nonanalytic corrections to scaling. 
Indeed, in $N$-vector systems, there are corrections $t^\Delta$ ($t$ is 
the reduced temperature) to the leading scaling behavior, 
with $\Delta\approx0.5$--$0.6$ in the physically relevant cases
$0\le N\le 4$. In the analysis of the MC data and of HT series, 
these slowly decaying
corrections require careful extrapolations, in the absence of which 
precise but incorrect results are obtained. To give an example,
we report here some old (but not too old!) 
results for the four-point renormalized coupling 
$g^*$ in the three-dimensional Ising model (see the discussion in 
Sec. 5 of Ref. \cite{PV-98} and Fig. 1 reported there):
\begin{table}[h]
\begin{center}
\begin{tabular}{ll}
MC, no nonanalytic corrections: \cite{BK-96}    & $g^* = 25.0(5)$; \\
MC, with  nonanalytic corrections: \cite{PV-98} & $g^* = 23.7(2)$; \\
HT, no nonanalytic corrections: \cite{ZLF-96}   & $g^* = 24.5(2)$; \\
HT, with  nonanalytic corrections: \cite{BC-97,PV-98} & 
                    $g^* = 23.69(10)$, $23.55(15)$.
\end{tabular}
\end{center}
\end{table}

\vskip - 0.4truecm 

\noindent
For comparison, perturbative field theory gives \cite{GZ-98} 
$g^* = 23.64(7)$, while a recent analysis of improved HT 
expansions gives \cite{CPRV-99} $g^* = 23.49(4)$.
Clearly, neglecting the corrections to scaling 
introduces a large systematic error! And, even worse, there is no way
to evaluate it, unless one assumes that nonanalytic corrections
are really there. 

A solution to these problems is represented by the so-called
improved models, \cite{CFN-82,HPV-99,BFMMPR-99,Hasenbusch-99,HT-99,CPRV-99,%
CPRV-00,CHPRV} which are such that the leading scaling correction
(approximately) vanishes. The systematic errors are now sensibly reduced
and one obtains more reliable estimates. 

MC and HT analyses, although different in practice,
are very similar in spirit, and indeed they are affected by the 
same type of systematic errors. In order to assess the reliability of the 
results, it is thus important to have a 
completely different approach to compare with. 
Field theory provides it and indeed 
independent estimates can be obtained using a variety of different
methods: the $\epsilon$-expansion pioneered by Wilson and
Fisher,
the fixed-dimension expansion proposed by Parisi, the perturbative 
expansion in the minimal-subtraction scheme without $\epsilon$-expansion
proposed by Dohm, and the so-called exact RG, which essentially
consists (there are many different versions, see, e.g., Ref. \cite{BB-review})
in {\em approximately} solving nonperturbatively the RG
equations. 

In this contribution we will focus on the fixed-dimension expansion
method, which 
is the one which has provided the most precise estimates, and, 
together with the $\epsilon$-expansion, has been the most used.
We consider the standard $\phi^4$ theory with $N$-vector fields
and discuss the role of the singularities of the RG functions 
at the critical point in the numerical determination of critical
quantities.

\vskip -0.5truecm

\section{Singularities of the RG functions}

An important controversial 
issue \cite{Parisi-80,NickelCS,Sokal94,Li-Madras-Sokal,Bagnuls97}
in the field-theoretic (FT) approach
in fixed dimension is the presence of nonanalyticities
at the fixed point $g^{*}$, which is defined as the zero of 
the $\beta$-function. The question was clarified
long ago by Nickel \cite{NickelCS} who gave a simple
argument to show that nonanalytic terms should in principle be present
in the $\beta$-function. The same argument applies also to other series,
like those defining the critical exponents:
Any RG function is expected to be singular at the fixed point.

To understand the problem, let us consider the four-point
renormalized coupling $g$ as a function of the temperature $T$. For
$T\to T_c$ we can write down an expansion of the form
\begin{eqnarray}
g = g^{*} &&
         \Bigl[ 1 + a_1 t + a_2 t^2 + \ldots +
                b_1 t^{\Delta} + b_2 t^{2 \Delta} + \ldots + \nonumber \\
&&  c_1 t^{\Delta+1} + \ldots +
    d_1 t^{\Delta_2} + \ldots +
    e_1 t^\gamma + \ldots \Bigr]\; ,
\label{grintermsTmTc}
\end{eqnarray}
where $\Delta$, $\Delta_2$, $\ldots$ are subleading exponents and 
$t$ is the reduced temperature $t \equiv (T-T_c)/T_c$.
The corrections proportional to $t^\gamma$ are due to the presence 
of an analytic background in the free energy.
We expect on general grounds that $a_1 = a_2 = a_3 = \ldots = 0$.
Indeed, these analytic corrections arise from the nonlinearity
of the scaling fields and their effect can be
eliminated in the Green's functions by an appropriate change of 
variables. \cite{Fisher-Aharony-analytic} For dimensionless
RG-invariant quantities such as $g$,
the leading term is universal and therefore independent of
the scaling fields, so that no analytic term can be generated. 
Analytic correction factors to the singular
correction terms are generally present, and
therefore the constants $c_i$ in
Eq.~(\ref{grintermsTmTc}) are expected to be nonzero. 

Starting from
Eq. (\ref{grintermsTmTc}) it is easy to compute the $\beta$-function.
Since the mass gap $m$ scales analogously, for $\Delta< \gamma$ (this
condition is usually, but not always, satisfied\footnote{
In some models, for instance in the 2D Ising model, $\Delta>\gamma$.
In this case, Eq. (\ref{betafunction}) is still correct \cite{CCCPV-00}
if $\gamma$ and $\Delta$ are interchanged. Also $\alpha_1 = - \gamma/\nu$
in this case.}),
we obtain the following expansion:
\begin{eqnarray}
\beta(g) = m {\partial g\over \partial m} &=& 
   \alpha_1 \Delta g + \alpha_2 (\Delta g)^2 +
               \ldots +
        \beta_1 (\Delta g)^{1\over \Delta} +
        \beta_2 (\Delta g)^{2\over \Delta} + \ldots + \nonumber \\
     && \gamma_1 (\Delta g)^{1+{1\over \Delta}} + \ldots +
        \delta_1 (\Delta g)^{\Delta_2\over \Delta} + \ldots
        +\zeta_1 (\Delta g)^{\gamma\over \Delta} + \ldots,
\label{betafunction}
\end{eqnarray}
where $\Delta g = g^* - g$.
It is easy to verify the well-known fact that
$\alpha_1 = -\Delta/\nu \equiv -\omega$ and that, if
$a_1 = a_2 = \ldots = 0$ in Eq. (\ref{grintermsTmTc}), then
$\beta_1 = \beta_2 = \ldots = 0$.
Eq. (\ref{betafunction}) clearly shows the presence of several 
nonanalytic terms with exponents depending on $1/\Delta$, $\Delta_i/\Delta$,
and $\gamma/\Delta$. 

As pointed out by Alan Sokal, \cite{Sokal94,Li-Madras-Sokal}
the nonanalyticity of the RG functions can also be understood within
Wilson's RG approach. We repeat here his argument. Consider the 
Gaussian fixed point which, for $3\le d < 4$, has a two-dimensional
unstable manifold ${\cal M}_u$: The two unstable directions correspond to the 
interactions $\phi^2$ and $\phi^4$. Then, notice that continuum
field theories are in a one-to-one correspondence with Hamiltonians
on ${\cal M}_u$ and that the FT RG is nothing but 
Wilson's RG restricted to ${\cal M}_u$. But now ${\cal M}_u$ has no 
special status at the nontrivial fixed point. In particular, there is no
reason why it should approach it along the leading irrelevant direction. 
Barring miracles, the approach should be along a generic direction 
which has nonzero components along any of the irrelevant directions.
If this happens, nonanalytic terms are present in any RG  function.

In order to clarify the issue, Ref. \cite{PV-98} determined the 
asymptotic behavior of $\beta(g)$
for $g\to g^*$ in the continuum field theory for 
$N\to\infty$ and $2<d<4$, and showed that Eq. (\ref{betafunction})
holds and that the expected nonanalytic terms are indeed present. 
In Ref. \cite{CCCPV-00} the computation was extended to two dimensions,
finding again nonanalytic terms. 

The presence of nonanalyticities gives rise to systematic deviations 
in FT estimates, as it did in MC and HT studies. 
In the next Section we will present a test-case and we will discuss 
the type of deviations one should expect.

\begin{figure}[t]
\centerline{\psfig{width=9truecm,angle=-90,file=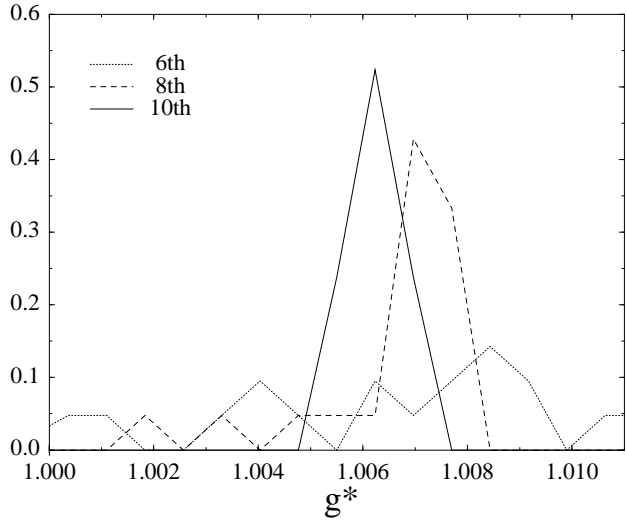}}
\caption{Distribution of the results for the resummations of $g^*$
for $p = 1/10$, $c = - Z_1/5$. 
\label{fig:gstar}}
\end{figure}

\section{A simple example}

In order to understand the role of the nonanalytic terms in the 
analyses of the FT perturbative expansions we have considered a simple zero 
dimensional example. Define
\be
f(g;c,p) = c (1-g)^{1+p} + \int_{-\infty}^\infty {dx\over \sqrt{2\pi}}
\; x^4\; \exp\left[- {x^2\over2} - {g x^4\over 24}\right].
\ee
For $c\not=0$ and $p$ not integer, this function has a branch point
for $g=1$ and thus it should mimic the behaviour we expect for FT
expansions. For $g\to 1$, we have 
\be
f(g;c,p) \approx Z_0 - Z_1 (1 - g) + c (1-g)^{1+p} + O((1-g)^2)),
\ee
where $Z_0 = 1.37556014$, $Z_1 = - 0.679325$. We wish to repeat here
the same steps performed in the calculation of $g^*$ and $\omega$. 
Therefore, we determine $g^*$ and $Z_1$, by solving the equations:
\be
   f(g^*;c,p) = Z_0, \qquad\qquad Z_1 = f'(g^*;c,p).
\ee
Of course, $f(g;c,p)$ is replaced with an appropriate resummation
of the its perturbative expansion. We use here the resummation scheme 
proposed in Ref. \cite{L-Z-77} that makes explicit use of the location of the 
Borel-transform singularity, but similar results are obtained 
extending the Borel transform by means of Pad\'e approximants
(we mention that one could also use the perturbative series in the 
{\em bare} coupling \cite{Kleinert_resummations}).
The mean values and errors are determined by using the procedure of 
Ref. \cite{CPV-00}. In the absence of nonanalytic term, i.e. for $c=0$,
using the $n$th-order expansion, we obtain
\begin{table}[h]
\begin{center}
\begin{tabular}{lll}
$n$ = 6:  & $g^* = 1.00025(131)$ & $Z_1 = - 0.6791(178)$; \\
$n$ = 8:  & $g^* = 0.99997(10)$  & $Z_1 = - 0.6800(18)$;  \\
$n$ = 10: & $g^* = 1.00000(1)$   & $Z_1 = - 0.6791(2) $.
\end{tabular}
\end{center}
\end{table}

\vskip -0.5truecm

\noindent There is good agreement, the precision increases by a factor
of 10 every two orders, and the error bars are correct. 

Then, we consider the role of the nonanalytic corrections, by adding 
a term that is small compared to the analytic one. 
We choose $c = -Z_1/5$. 

\begin{figure}[t]
\centerline{\psfig{width=9truecm,angle=-90,file=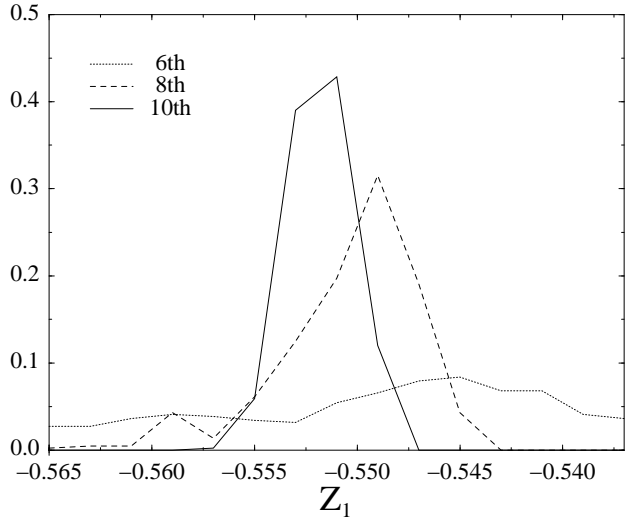}}
\caption{Distribution of the results for the resummations of $Z_1$
for $p = 1/10$, $c = - Z_1/5$. 
\label{fig:omega}}
\end{figure}

Now, for $p=1/10$ we obtain
\begin{table}[h]
\begin{center}
\begin{tabular}{lll}
$n$ = 6:  & $g^* = 1.0043(62)$  & $Z_1 = - 0.550(20)$; \\
$n$ = 8:  & $g^* = 1.0066(15)$  & $Z_1 = - 0.550(3)$;  \\
$n$ = 10: & $g^* = 1.0062(5)$   & $Z_1 = - 0.552(2)$.
\end{tabular}
\end{center}
\end{table}

\vskip -0.4truecm

\noindent
In this case
the agreement is poor, especially for $Z_1$, and, even worse,
the errors are completely incorrect. This can be understood from 
Figs. \ref{fig:gstar} and \ref{fig:omega} where we show the distribution of the 
approximants that are used. These distributions are nicely peaked, but
unfortunately at an incorrect value of $g^*$ and $Z_1$. 
Thus, in the presence of these (strong) nonanalyticities, the 
fact that the approximants  have a narrow distribution is not
a good indication that the results are reliable.
Also, the stability of the results with the number of terms of the series 
is completely misleading. As we shall discuss below, this is what we believe
is happening in two dimensions.

If we instead consider a weak nonanalyticity, i.e. $p\approx 1$,
the discrepancies we have found for $p=1/10$ 
are much smaller, although still present. For instance, for $p = 9/10$
\begin{table}[h]
\begin{center}
\begin{tabular}{lll}
$n$ = 6:  & $g^* = 1.013(45)$  & $Z_1 = - 0.755(270)$; \\
$n$ = 8:  & $g^* = 1.014(7)$   & $Z_1 = - 0.655(23)$;  \\
$n$ = 10: & $g^* = 1.006(4)$   & $Z_1 = - 0.668(2) $. 
\end{tabular}
\end{center}
\end{table}

\vskip -0.4truecm

\noindent
In this case the results are consistent with the exact values, although
the errors are still slightly underestimated. As expected, the 
largest discrepancies are observed for $Z_1$.

\section{Conclusions}

In the previous Section we have shown that nonanalytic terms may give rise 
to systematic deviations and a systematic underestimate of the error bars.
Now, what should we expect in the interesting two- and three-dimensional 
cases? 

In three dimensions $\Delta\approx 0.5$ and $\Delta_2/\Delta$ 
is approximately \cite{Riedel} 2. Thus, the leading nonanalytic term
has exponent $\Delta_2/\Delta$ and is not very different from an analytic one.
Thus, we expect small corrections and indeed the FT results 
are in substantial agreement with the estimates obtained in
MC and HT studies. However, small differences are observed for $\gamma$ and 
$\omega$ for $N=0$ and $N=1$. For instance $(\Delta = \omega\nu)$

\begin{table}[h]
\begin{center}
\begin{tabular}{lllll}
$N=0$ & $\gamma_{\rm FT} = 1.1596(20)$ & Ref. \cite{GZ-98}, & \qquad
        $\gamma_{\rm MC} = 1.1575(6)$  & Ref. \cite{C-C-P}, \\
$N=0$ & $\Delta_{\rm FT} = 0.478(10)$  & Ref. \cite{GZ-98}, & \qquad
        $\Delta_{\rm MC} = 0.517(7)^{+10}_{-0}$ & Ref. \cite{B-N-97}, \\
$N=1$ & $\gamma_{\rm FT} = 1.2396(13)$ & Ref. \cite{GZ-98}, & \qquad
        $\gamma_{\rm HT} = 1.2371(4)$  & Ref. \cite{CPRV-99}, \\
$N=1$ & $\Delta_{\rm FT} = 0.504(8)$   & Ref. \cite{GZ-98}, & \qquad
        $\Delta_{\rm MC} = 0.533(6)$    & Ref. \cite{Habilitation}.
\end{tabular}
\end{center}
\end{table}

\vskip -0.5truecm

\noindent There are slight differences, especially for $\omega$, but still
at the level of a few error bars. Note that, as discussed in Ref. \cite{PV-98},
part of the error may be due to a slightly incorrect estimate of $g^*$.
Using the estimate of $g^*$ obtained from the analysis of the HT expansions,
the FT estimates change towards the HT and MC values.

Larger discrepancies are observed in two dimensions. 
For $N\ge 3$ it is easy to predict the behavior of the RG functions at
the critical point. Indeed, the theory is massive
for all temperatures. The critical behavior is controlled by the
zero-temperature Gaussian point and can be studied in perturbation theory
in the corresponding $N$-vector model. One finds only logarithmic
corrections to the purely Gaussian behavior.
It follows that the operators have dimensions that coincide
with their naive (engineering)  dimensions,
apart from logarithmic multiplicative
corrections related to the so-called anomalous dimensions.
The leading irrelevant operator has dimension two \cite{BLZ,Symanzik} and thus,
for $m\to 0$, we expect 
\begin{equation}
g(m) = g^* \left\{ 1 + c \,m^2 \left( - \ln m^2\right)^\zeta
\left[1 + O\left( {\ln (-\ln m^2)\over \ln m^2}\right)\right]\right\},
\label{gmsc}
\end{equation}
where $\zeta$ is an exponent related to the anomalous dimension of the
leading irrelevant operator, and $c$ is a constant. Therefore,
\begin{equation}
\beta(g) = m {\partial g\over \partial m}= - 2 \Delta g
   \left( 1 + {\zeta\over \ln \Delta g} + \cdots \right),
\label{bgn3}
\end{equation}
with $\Delta g\equiv g^*-g$. Clearly there are in this case logarithmic 
corrections and therefore, we expect large deviations in the determinations
of $\omega$, which should be 2. These deviations are indeed observed:
for $N=3$ the analysis of the five-loop series \cite{OS-00}
yields the estimate $\beta'(g^*) = 1.33(2)$, which is very different
from the expected result $\beta'(g^*) = 2$.

For $N=1$, we can repeat Nickel's analysis, using the fact 
that, from conformal field theory, we can compute the RG dimensions of 
all relevant and  irrelevant operators. Using these results 
we predict that\footnote{It has been claimed sometimes that 
in the Ising universality class the leading irrelevant operator
has $\omega=4/3$, so that $\beta'(g^*) =4/3$. This claim is incorrect. 
Indeed, such operator can only appear in nonunitary extensions of the 
Ising model, but not in the standard (unitary) $\phi^4$ field theory. 
For a detailed 
discussion, see App. A of Ref. \cite{CCCPV-00}.}
\begin{equation}
\beta(g) =
 - {7\over 4} \Delta g
\left( 1 + b_1 |\Delta g|^{1/7}+ b_2 |\Delta g|^{2/7} + b_3 |\Delta g|^{3/7}
+   \cdots \right)
\label{nan}
\end{equation}
where $\Delta g\equiv g^*-g$. Such an expansion is confirmed by an analysis 
of the lattice Ising model. Again we find strong nonanalyticities, and 
correspondingly we expect large deviations. And, indeed such large deviations
are observed: The analysis of the five-loop series gives \cite{OS-00} 
$g^*=15.39(25)$ and $\beta'(g^*) = 1.31(3)$, to be compared with 
the exact prediction $\beta'(g^*) = 7/4$ and 
the estimates $g^*=14.69735(3)$ (Ref. \cite{CHPV-g4}) and 
$g^* = 14.6975(1)$ (Ref. \cite{BNNPSW-00}).



\def\PLA{{\em Phys. Lett.} A}
\def\PRB{{\em Phys. Rev.} B}
\def\PRD{{\em Phys. Rev.} D}
\def\PRE{{\em Phys. Rev.} E}
\def\PRL{{\em Phys. Rev.Lett.}}
\def\NPB{{\em Nucl. Phys.} B}
\def\PLB{{\em Phys. Lett.} B}
\def\JPA{{\em J. Phys.} A}
\def\JSP{{\em J. Stat. Phys.}}
\def\EE{{\em Europhys. Lett.}}
\def\JPhysStud{{\em J. Phys. Stud.}}
\def\FTT{{\em Fiz. Tverd. Tela}}

\end{document}